\documentclass{amsart}

\usepackage[fontsize=12pt]{scrextend} 
\usepackage[english]{babel}
\usepackage{amssymb,amsmath,amsthm}
\usepackage{graphicx}
\usepackage{color}
\usepackage{dsfont}
\usepackage[font=small, width=\textwidth]{caption}

\newcommand{\Cov}{\operatorname{Cov}}

\theoremstyle{plain}
\theoremstyle{remark}
\theoremstyle{plain}
\theoremstyle{plain}
\theoremstyle{plain}
\theoremstyle{plain}
\theoremstyle{plain}
\theoremstyle{remark}

\title[Block Length Choice for the Bootstrap of Functional Data]{Block Length Choice for the Bootstrap of Dependent Panel Data \\ \small - a Comment on Choi and Shin (2020)}

\author[Wegner]{Lea Wegner}

\author[Wendler]{Martin Wendler}
\thanks{The research was supported by the German Research Foundation (Deutsche Forschungsgemeinschaft – DFG), project WE 5988/3 \emph{Analyse funktionaler Daten ohne Dimensionsreduktion}.}
\address{Otto-von-Guericke-Universit\"at Magdeburg, Germany}
\email{martin.wendler@ovgu.de}
\date{\today}


\begin{document}

\begin{abstract} Choi and Shin \cite{choi2020} have constructed a bootstrap-based test for change-points in panels with temporal and and/or cross-sectional dependence. They have compared their test to several other proposed tests. We demonstrate that by an appropriate, data-adaptive choice of the block length, the change-point test by Sharipov, Tewes, Wendler \cite{sharipov2016} can at least cope with mild temporal dependence, the size distortion of this test is not as severe as claimed by Choi and Shin \cite{choi2020}.

\end{abstract}

\maketitle

\section{Introduction}\label{sec1:intro}

Choi and Shin \cite{choi2020} have suggested a change-point test for panel data of the form $(X_{i,t})_{i=1,..,N,t=1,...,T}$, and allow possible temporal correlation (that means $\Cov(X_{i,t},X_{i,s})\neq 0$ for $s\neq t$) or cross sectional correlation ($\Cov(X_{i,t},X_{j,u})\neq 0$ for $i\neq j$). They combine a test statistic introduced by Horv{\'a}th and Hu{\v{s}}kov{\'a} \cite{horvath2012} once with circular bootstrap introduced by Politis and Romano \cite{politis1992} and once with the stationary bootstrap \cite{politis1994}. The test statistic used by Choi and Shin is given by
\begin{equation*}
H_{N,T}=\max_{t=1,...,T-1}\frac{1}{\sqrt{N}}\sum_{i=1}^N\bigg(\frac{1}{\hat{\sigma_i}^2T}\Big(\sum_{r=1}^t(X_{i,r}-\bar{X}_i)\Big)^2-\frac{t(T-t)}{T^2}\bigg),
\end{equation*}
where $\bar{X}_i=\sum_{t=1}^T X_{i,t}/T$ and $\hat{\sigma_i}^2$ is the Bartlett estimator of the variance for the partial sum. An alternative test statistic introduced by Sharipov, Tewes and Wendler \cite{sharipov2016} could be replaced by the equivalent statistic
\begin{equation*}
J^2_{N,T}=\max_{t=1,...,T-1}\sum_{i=1}^N\frac{1}{T}\Big(\sum_{r=1}^t(X_{i,r}-\bar{X}_i)\Big)^2.
\end{equation*}
The authors obtain critical values by using the non-overlapping block bootstrap (Carlstein \cite{carlstein1986}). So the main difference between the two approaches is the estimator $\hat{\sigma_i}^2$ of the long-run variance. It thus can be expected that the test by Choi and Shin performs better under pronounced temporal dependence, and this is confirmed by their simulation study. However, we will demonstrate that the performance of the test by Sharipov, Tewes, Wendler is much better under mild dependence than claimed by Choi and Shin, if an appropriate, data-adaptive block length is chosen.

\section{Block Length Choice}\label{sec2:block}

The problem of choosing a block length is related to the choice of the bandwidth in a kernel variance estimation. The reason is that the variance of the bootstrap partial sum is similar to the Bartlett estimator. Rice and Shang \cite{rice2017} have proposed a bandwidth selection method for the long-run variance estimation of functional time series. We adopt their method for the block bootstrap. The procedure is the following: \\
\begin{itemize}
\item determine a starting value $L_0=T^{1/2}$
\item calculate matrices

$V_k = \frac{1}{T}\sum_{t=1}^{T-(k-1)} (X_{1,t},...,X_{N,t}) \otimes (X_{1,(t+k)},...,X_{N,(t+k)}) $ for $k=1,..., L_0$, where $\otimes$ is the outer product
\item compute
\begin{align*}
CP_0&=V_1+2\sum_{k=1}^{L_0-1}w(k, L_0)V_{k+1}\\
\text{and} \quad CP_1&= 2\sum_{k=1}^{L_0-1}k \,w(k, L_0)V_{k+1}
\end{align*}
where in our case $w$ is Bartlett kernel  $w(k,L)= (1-|\frac{k}{L}|)\mathbf{1}_{\{|\frac{k}{L}| \leq 1 \}}$
 \item Receive the data adapted bandwidth 
 \begin{equation*}
 L_{adpt} = \left \lceil \left( \frac{3T\sum_{i=1}^N \sum_{j=1}^N {CP_1}_{i,j}}{\sum_{i=1}^N\sum_{j=1}^N{CP_0}_{i,j}+ \sum_{j=1}^N {CP_0}_{j,j}^2 }  \right)^{1/5}    \right \rceil   
 \end{equation*}
\end{itemize}
Note that the block length has to be an integer, so we have to round up the value obtained from the procedure by Rice and Shang \cite{rice2017}.

\section{Simulation Study}\label{sec3:simus}

Following \cite{choi2020}, we consider the following structure for data simulation:
\begin{equation*}
X_{i,t} = \delta_i \mathbf{I}_{\{t>t_0\}} +e_{i,t}, \; e_{i,t}= \rho_i e_{i,t-1}+\varepsilon_{i,t}, \; \varepsilon_{i,t}= a_{i,t}+\beta_i f_t
\end{equation*}
for $i=1,...,N, \; t=1,...,T$, where $\mathbf{I}$ is the indicator function and where $a_{i,t}$ and $f_t$ ($i=1,...,N \; t=1,...,T$) are independent zero mean errors having either the standard normal distribution or the standardized $t$-distribution with 5 degrees of freedom. Under the hypothesis of no change, we have $\delta_i=0$ for all $i=1,...,N$, while under the alternative, we generate the $\delta_i$ at random. The parameters $\rho_i$ and $\beta_i$ model the serial and the cross-sectional dependence. For the serial correlation parameter the values $\rho_i\in \{0,\ 0.3,\ 0.5\}$ are considered. The cross-sectional correlation parameter is chosen as $\beta_i \in \{0,\  0.5, \ 2\}$.

Table \ref{table size} shows the empirical rejection frequencies under the hypothesis of no change (all $\delta_i=0$) for $S=1000$ simulation runs. $H^{\star CB}$ and $H^{\star SB}$ denote the test of Horv{\'a}th and Hu{\v{s}}kov{\'a} \cite{horvath2012} combined with the circular or the stationary bootstrap as implemented by Choi and Shin \cite{choi2020}. $J^{\star CS}$ denotes the test based on $J_{N,T}$ and nonoverlapping block bootstrap with the block length choice as proposed by Choi and Shin \cite{choi2020}, while $J^{\star RS}$ uses the same test statistic, but with the block length choice using the method by Rice and Shang \cite{rice2017} as described in the previous section.

\begin{table}[!htbp]
\centering
\resizebox{\columnwidth}{!}{%
\begin{tabular}{ |c c c c | c c c c | c c c c| } 
\hline
\multicolumn{12}{|c|}{Empirical size of level $\alpha = 0.05$ tests} \\
\hline
       &         &     &     &  \multicolumn{4}{|c|}{$\mathcal{N}(0,1)$ error terms} & \multicolumn{4}{|c|}{$t_5$ error terms }\\
$\rho_i$ & $\beta_i$ & $N$ & $T$ & $H^{\star CB}$ & $H^{\star SB}$ & $J^{\star CS}$ & $J^{\star RS}$ &  $H^{\star CB}$ & $H^{\star SB}$ & $J^{\star CS}$ & $J^{\star RS}$ \\
\hline
$0.3$ & $0$ & $50$ & $50$  & $0.057$ & $0.107$ & $0.216$ & $0.045$ & $0.077$ & $0.134$ & $0.237$ & $0.075$ \\
 	  & 	& $50$ & $100$ & $0.029$ & $0.056$ & $0.252$ & $0.033$ & $0.050$ & $0.074$ & $0.230$ & $0.032$ \\
 	  & 	& $100$ & $50$ & $0.087$ & $0.127$ & $0.257$ & $0.009$ & $0.078$ & $0.119$ & $0.252$ & $0.007$ \\
 	  & 	& $100$ & $100$ & $0.037$ & $0.061$ & $0.254$ & $0.006$ & $0.033$ & $0.058$ & $0.242$ &$0.007$ \\
 	  & 	& $200$ & $100$ & $0.035$ & $0.062$ & $0.234$ &  $0$    & $0.032$ & $0.055$ & $0.249$ & $0$       \\ 	
 	  & 	& $100$ & $1000$ & $0.041$ & $0.050$ & $0.378$ & $0.059$ & $0.039$ & $0.040$ & $0.389$ &  $0.054$   \\ 	
\hline 	  
$0.5$ & $0$ & $50$ & $50$  & $0.040$ & $0.064$ & $0.447$ & $0.354$ & $0.038$ & $0.061$ & $0.438$ & $0.329$ \\
 	  & 	& $50$ & $100$ & $0.015$ & $0.028$ & $0.487$ & $0.234$ & $0.033$ & $0.047$ & $0.451$ & $0.175$ \\
 	  & 	& $100$ & $50$ & $0.036$ & $0.053$ & $0.454$ & $0.228$ & $0.040$ & $0.060$ & $0.465$ & $0.192$ \\
 	  & 	& $100$ & $100$ & $0.025$ & $0.043$ & $0.481$ & $0.122$ & $0.018$ & $0.033$ & $0.455$ & $0.108$ \\
 	  & 	& $200$ & $100$ & $0.021$ & $0.030$ & $0.473$ &  $0.070$ & $0.017$ & $0.028$ & $0.473$ & $0.045$ \\ 	
 	  & 	& $100$ & $1000$ & $0.041$ & $0.046$ & $0.659$& $0.272$ & $0.077$ & $0.016$ & $0.646$ & $0.279$   \\ 	
\hline 	    
$0$   & $0.5$ & $50$ & $50$  & $0.020$ & $0.025$ & $0.024$ & $0.048$ & $0.017$ & $0.020$ & $0.018$ & $0.038$ \\
 	  & 	& $50$ & $100$ & $0.034$ & $0.040$ & $0.032$ & $0.047$ & $0.029$ & $0.033$ & $0.032$ & $0.029$ \\
 	  & 	& $100$ & $50$ & $0.018$ & $0.021$ & $0.016$ & $0.038$ & $0.017$ & $0.023$ & $0.014$ & $0.030$ \\
 	  & 	& $100$ & $100$ & $0.025$ & $0.029$ & $0.026$ & $0.038$ & $0.022$ & $0.026$ & $0.023$ & $0.023$ \\
 	  & 	& $200$ & $100$ & $0.031$ & $0.035$ & $0.034$ &  $0.023$ & $0.039$ & $0.037$ & $0.023$ & $0.015$ \\ 	
 	  & 	& $100$ & $1000$ & $0.056$ & $0.058$ & $0.054$ &  $0.037$ & $0.058$ & $0.061$ & $0.056$ &  $0.028$  \\ 
\hline 	  	
$0$   & $2$ & $50$ & $50$  & $0.048$ & $0.044$ & $0.041$ & $0.113$ & $0.050$ & $0.055$ & $0.043$ & $0.068$ \\
 	  & 	& $50$ & $100$ & $0.035$ & $0.031$ & $0.035$ & $0.068$ & $0.043$ & $0.037$ & $0.033$ & $0.059$ \\
 	  & 	& $100$ & $50$ & $0.044$ & $0.046$ & $0.039$ & $0.063$ & $0.035$ & $0.038$ & $0.032$ & $0.063$ \\
 	  & 	& $100$ & $100$ & $0.058$ & $0.052$ & $0.054$ & $0.046$ & $0.051$ & $0.051$ & $0.043$ & $0.052$ \\
 	  & 	& $200$ & $100$ & $0.050$ & $0.039$ & $0.045$ &  $0.037$ & $0.050$ & $0.046$ & $0.045$ & $0.053$ \\ 	
 	  & 	& $100$ & $1000$ & $0.066$ & $0.060$ & $0.067$ & $0.043$ & $0.064$ & $0.059$ & $0.065$ &  $0.050$   \\  	
\hline 	      
$0.3$ & $0.5$ & $50$ & $50$  & $0.048$ & $0.077$ & $0.123$ & $0.139$ & $0.053$ & $0.077$ & $0.157$ & $0.111$ \\
 	  & 	  & $50$ & $100$ & $0.045$ & $0.054$ & $0.156$ & $0.100$ & $0.032$ & $0.043$ & $0.154$ & $0.083$ \\
 	  & 	  & $100$ & $50$ & $0.058$ & $0.074$ & $0.157$ & $0.091$ & $0.063$ & $0.084$ & $0.175$ & $0.072$ \\
 	  & 	  & $100$ & $100$ & $0.039$ & $0.050$ & $0.149$ & $0.058$ & $0.031$ & $0.037$ & $0.148$ & $0.068$ \\
 	  & 	  & $200$ & $100$ & $0.042$ & $0.052$ & $0.154$ & $0.061$ & $0.036$ & $0.047$ & $0.155$ & $0.065$ \\ 	
 	  & 	  & $100$ & $1000$ & $0.053$ & $0.058$ & $0.174$ & $0.086$ & $0.058$ & $0.051$ & $0.154$ & $0.069$  \\ 
\hline 	  
\end{tabular}%
}
\caption{Empirical rejection frequencies of the tests $H^{\star CB}$ and $H^{\star SB}$  based on $H_{N,T}$ and of the tests  $J^{\star CS}$ and $J^{\star RS}$ based on $J_{N,T}$ for different values of the correlation parameters ($\rho_i$, $\beta_i$) and of the sample size ($N$, $T$) (values for  $H^{\star CB}$, $H^{\star SB}$, $J^{\star CS}$ taken from \cite{choi2020}).} 
\label{table size}
\end{table}

As observed in \cite{choi2020}, $H^{\star CB}$ and $H^{\star SB}$  does not exceed the theoretical size of 5\% much under serial and under cross-sectional dependence. However, with the new data-adaptive block length choice, the test $J^{\star RS}$ is not oversized at least for mild serial dependence ($\rho=0.3$), and the size distortion for stronger serial dependence ($\rho=0.5$) is much less severe compared to $J^{\star CS}$. So it turns out that the claim of Choi and Shin \cite{choi2020} that the test $J^{\star}$ introduced in \cite{sharipov2016} has size distortion problem for serially correlated panels is in part due to a non-optimal implementation of the block bootstrap. However, we acknowledge that the tests $H^{\star CB}$ and $H^{\star SB}$ have better size properties under stronger serial dependence.

The empirical power of the different tests is shown in Table \ref{table power}. Here, only the case of normally distributed $a_{i,t}$ and $f_t$ is considered, and the correlation parameters are chosen as $\rho_i=\beta_i=0$ (no dependence). The size and direction of the change are choosen at random: The $\delta_i$ are independent, following two uniform distributions: $\mathcal{U}(-\frac{1}{2},\frac{1}{2})$ (cancelling break) or $\mathcal{U}(\frac{1}{10},\frac{1}{2})$ (non-cancelling break). For the time of the change we study $t_0=0.3T$ and $t_0=0.5T$. For the canceling breaks, the tests $H^{\star CB}$ and $H^{\star SB}$ have higher power, while for non-cancelling breaks, the tests $J^{\star CS}$ and $J^{\star RS}$ reject the hypothesis more often. Under the alternative (at least under independence), the power does not seem to be much influenced by the block length choice.

\begin{table}[!htbp]
\resizebox{\columnwidth}{!}{%
\begin{tabular}{ |c c c | c c c c | c c c c | } 
\hline
\multicolumn{11}{|c|}{Empirical power of level $\alpha=0.05$ tests} \\
\hline
      &     &     & \multicolumn{4}{|c|}{$\delta_i \sim \mathcal{U}(\frac{1}{2},\frac{1}{2})$} & \multicolumn{4}{|c|}{$\delta_i \sim \mathcal{U}(\frac{1}{10},\frac{1}{2})$ } \\
$t_0$ & $N$ & $T$ &  $H^{\star CB}$ & $H^{\star SB}$ & $J^{\star CS}$ & $J^{\star RS}$ &  $H^{\star CB}$ & $H^{\star SB}$ & $J^{\star CS}$ & $J^{\star RS}$ \\
\hline
$0.5T$ & $50$ & $50$ & $0.683$ & $0.818$ & $0.727$ & $0.506$ & $0.531$ & $0.687$ & $0.743$ & $0.738$ \\
       & $50$ & $100$ & $0.997$ & $1$ & $0.998$ & $0.987$ & $0.993$ & $0.997$ & $0.996$ & $1$ \\
       & $100$ & $50$ & $0.765$ & $0.921$ & $0.819$ & $0.631$ & $0.559$ & $0.733$ & $0.807$ & $0.868$  \\       
       & $100$ & $100$ & $0.998$ & $1$ & $1$ & $1$ & $0.999$ & $1$ & $0.999$ & $1$ \\  
       & $200$ & $100$ & $0.999$ & $1$ & $1$ & $1$   & $0.997$ & $1$ & $1$ & $1$     \\        
       & $100$ & $1000$ & $1$ & $1$ & $1$ &  $1$   & $1$  & $1$  & $1$ &   $1$  \\ 
\hline
$0.3T$ & $50$ & $50$ & $0.394$ & $0.507$ & $0.310$ & $0.173$ & $0.202$ & $0.316$ & $0.274$ & $0.340$  \\
       & $50$ & $100$ & $0.990$ & $1$ & $0.958$ & $0.745$ & $0.933$ & $0.983$ & $0.913$ & $0.915$ \\
       & $100$ & $50$ & $0.428$ & $0.592$ & $0.308$ & $0.090$ & $0.217$ & $0.336$ & $0.323$ & $0.281$ \\       
       & $100$ & $100$ & $0.995$ & $0.999$ & $0.966$ & $0.933$ & $0.996$ & $1$ & $0.992$ & $0.993$\\  
       & $200$ & $100$ & $0.991$ & $1$ & $0.957$ & $0.995$ & $0.993$ & $1$ & $0.994$ & $1$ \\        
       & $100$ & $1000$ & $1$ & $1$ & $1$ & $1$ & $1$ & $1$ & $1$ &     $1$ \\ 
\hline
                
\end{tabular}%
}
\caption{Empirical rejection frequencies of the tests $H^{\star CB}$ and $H^{\star SB}$  based on $H_{N,T}$ and of the tests  $J^{\star CS}$ and $J^{\star RS}$ based on $J_{N,T}$ for different values of the correlation parameters ($\rho_i$, $\beta_i$) and of the sample size ($N$, $T$) (values for  $H^{\star CB}$, $H^{\star SB}$, $J^{\star CS}$ taken from \cite{choi2020}).}
\label{table power} 
\end{table}

\end{document}